# Convenience vs. Control: A Qualitative Study of Youth Privacy with Smart Voice Assistants

Molly Campbell
*Computer Science Department*
*Vancouver Island University*
Nanaiomo, BC, Canada
molly.campbell@viu.ca

Trevor De Clark
*Computer Science Department*
*Vancouver Island University*
Nanaiomo, BC, Canada
trevor.declark@viu.ca

Mohamad Sheikho Al Jasem
*Computer Science Department*
*Vancouver Island University*
Nanaiomo, BC, Canada
mohamad.sheikhoaljasem @viu.ca

Sandhya Joshi
*VIU Affiliate*
*Vancouver Island University*
Nanaiomo, BC, Canada
sandhya.joshi@viu.ca

Ajay Kumar Shrestha
*Computer Science Department*
*Vancouver Island University*
Nanaiomo, BC, Canada
ajay.shrestha@viu.ca

***Abstract*— Smart voice assistants (SVAs) are embedded in the daily lives of youth, yet their privacy controls often remain opaque and difficult to manage. Through five semi-structured focus groups (N=26) with young Canadians (ages 16-24), we investigate how perceived privacy risks (PPR) and benefits (PPBf) intersect with algorithmic transparency and trust (ATT) and privacy self-efficacy (PSE) to shape privacy-protective behaviors (PPB). Our analysis reveals that policy overload, fragmented settings, and unclear data retention undermine self-efficacy and discourage protective actions. Conversely, simple transparency cues were associated with greater confidence without diminishing the utility of hands-free tasks and entertainment. We synthesize these findings into a qualitative model in which transparency friction erodes PSE, which in turn weakens PPB. From this model, we derive actionable design guidance for SVAs, including a unified privacy hub, plain-language "data nutrition" labels, clear retention defaults, and device-conditional micro-tutorials. This work foregrounds youth perspectives and offers a path for SVA governance and design that empowers young digital citizens while preserving convenience.**

***Keywords*— *Privacy, Smart Devices, Focus Group, Youth, AI, Voice Assistants, User control, Qualitative Analysis***

## I. INTRODUCTION

Smart Voice Assistants (SVAs) have seamlessly integrated into the daily lives of youth with the promise of hands-free convenience as they navigate the complex environments of home, school, and public settings. However, this convenience is shadowed by persistent privacy questions. Young users are often at the forefront of adopting these technologies [1], [2], [3], yet they operate in a landscape where data collection is continuous, often opaque, and embedded in the background of their everyday routines [4], [5].

Existing frameworks, such as the privacy calculus model and the concept of the privacy paradox, provide a foundational understanding of the trade-offs users make between the perceived benefits and the perceived risks of technology use [6], [7], [8], [9]. However, these broad perspectives often leave the lived experiences of youth underspecified. Much of the existing privacy research is centered on adult populations, whose priorities, technological fluency, and social contexts differ significantly from those of young people, a gap highlighted by the lack of youth-specific measurement tools and frameworks [1], [10]. Consequently, the concrete points where transparency breaks down, where feelings of control are lost, and how these failures directly affect a young person's protective behavior remain critical gaps in the literature.

This study employs a qualitative approach through five semi-structured focus-group discussions (N=26) to investigate the nuanced experiences of Canadian youth. Our central thesis is that youth negotiate SVA use as a dynamic convenience-risk tradeoff between perceived privacy risks (PPR) and perceived benefits (PPBf). We argue that opaque algorithmic practices from manufacturers depress privacy self-efficacy (PSE), which in turn constrains privacy-protective behaviors (PPB). The process is mediated by the youth's perception of algorithmic transparency and trust (ATT) at key interaction points. We argue that enhancing transparency at critical action points, such as in settings, activity histories, and moments of consent, can improve PSE and PPB without sacrificing utility. By focusing on these five key constructs (PPR, PPBf, ATT, PSE, PPB), this study maps the pathway from perception to action in youths' privacy management of SVAs. Furthermore, by centering youth voices, this research moves beyond viewing them as mere consumers to engaging them as digital citizens, thereby identifying the specific supports needed to enhance their PSE.

The contributions of this work bridge empirical findings with practical application. First, we present a construct-aligned qualitative taxonomy that provides a nuanced, youth-centered understanding of the key privacy constructs (PPR, PPBf, ATT, PSE, PPB). Second, we move beyond taxonomy to propose an empirically grounded qualitative model that illustrates how friction in ATT leads to low PSE and ultimately influences PPB, while also identifying concrete design levers, such as plain-language settings and default opt-outs, to disrupt this pathway. Finally, through a compliance-experience gap analysis, we triangulate youth narratives with findings from formal prior privacy audit evidence to reveal where policy and design compliance fail to meet the practical needs of young users [11]. These contributions culminate in actionable youth-friendly guidance and the release of anonymized research material to support replication and future research.

The remainder of the paper is structured as follows: Section II provides the background and related work. Section III details the methodology. Results are presented in Section IV, with the discussion and limitations following in Section V. Finally, Section VI concludes the paper.

## II. Background and Related Work

As digital technologies become increasingly embedded in everyday life, understanding how users balance convenience with privacy has gained significant attention in Human-Computer Interaction (HCI) and privacy research. The Privacy Calculus Model (PCM) provides a foundational lens, suggesting that individuals make deliberate decisions about sharing personal information by weighing perceived benefits against potential privacy risks [6]. These trade-offs are central in HCI inquiries that examine how interface design and interaction design can shape perceptions and influence behavior [12], [13]. A well-documented counterpoint is the Privacy Paradox, which demonstrates that users frequently express distrust toward data-driven systems yet continue to engage with them daily, illustrating a gap between privacy attitudes and actual behaviors [8], [14]. This phenomenon has been widely documented across various digital contexts, including SVAs. A primary disruptor is a lack of transparency and trust. The opacity of data flows in Internet-of-Things (IoT) ecosystems, specifically the always-listening, wake-word detection of SVAs, creates uncertainty for users about what data is being collected, when processing occurs locally, and when data are transmitted to the cloud services [15], [16]. From an HCI perspective, transparency is not only a disclosure problem but an interaction design problem: information about data practices must be presented in ways that are understandable, timely, and actionable for end users [17].

Within this broader landscape, youth constitute a particularly important and, at times, vulnerable population. Centering youth voices in privacy research ensures that their perspectives are meaningfully represented. It also helps identify gaps in education and highlights how young people can be better equipped to make informed decisions about data privacy and consent. Rather than viewing youth merely as consumers, recent work encourages engaging them as digital citizens capable of questioning, shaping and imagining alternative futures for technology [18]. Empowering youth to better understand and act on their privacy rights supports this development [10]. Youth-centered approaches to consent and transparency are particularly impactful because younger users may face additional barriers in interpreting data flows and privacy policies. Most youth report learning about artificial intelligence (AI) and data practices primarily through popular culture, such as science fiction, movies, TV, and video games, which shows that formal education is often insufficient [19]. As knowledge and information processing skills evolve with age, youth benefit from supports that foster a basic, actionable understanding of how data are handled [20].

Regulatory and legal frameworks, such as Canada's Personal Information Protection and Electronic Documents Act (PIPEDA) [21], and design frameworks, such as Privacy-by-Design (PbD) principles [5], provide crucial guidance by mandating informed consent [22] and embedding privacy into systems, respectively. However, these frameworks are often designed with a generic adult user in mind, focusing on legal compliance instead of practical usability [5]. Consequently, despite these safeguards, users, and youth in particular, often remain uncertain about the lifecycle of their personal information, as the controls provided are often complex, hidden or incomprehensible [4].

This study addresses a critical gap at the intersection of these strands. Prior work has established the importance of the privacy calculus, the paradox of user behavior, and the need for transparency. Yet the specific mechanisms by which PPR, PPBf, ATT, and PSE translate into PPB for SVA use in youth remain underexplored [23]. Existing adult-centric models and compliance-focused frameworks alone do not adequately explain the precise mechanisms that connect young users' perception of risk and benefit to their privacy behaviors. This fragmented research landscape is summarized in Table I. As the table illustrates, prior work frequently examines isolated constructs, employs methods that lack a youth focus, and fails to provide actionable design levers. Our study addresses this gap by examining how Canadian youth manage privacy with SVAs, to move beyond theory to identify design and policy strategies that improve transparency, strengthen self-efficacy and support protective behavior without sacrificing convenience.

TABLE I. Related Work-Map

| Population | Method | Construct Coverage | Reported Gaps |
|---|---|---|---|
| Mixed [4] | Legal Analysis | PPB, ATT | Limited youth focus; Compliance-centric orientation |
| Adult [5] | Conceptual Analysis, Technical Overview, Literature Review | PPR, ATT | Lacks empirical focus on PPB and PSE; Limited youth focus; Focus on system rather than user privacy |
| Youth [6] | Surveys, Interviews, Focus Groups | PPR, PPBf, ATT | Broad stakeholder perceptions; Few actionable ATT or PSE design levers; Limited information on PPB after data collection |
| Adult [8] | Surveys, Ethnographic | PPR, ATT | Lacks youth perspective; No actionable design levers; Short-term focus |
| Youth [10] | Surveys | PSE | Artificial tasks; Lacks empirical focus on PPB; Few actionable PSE or ATT design levers |
| Mixed [14] | Surveys | PPR, PPBf, ATT | Focuses on behavioral intention, not PPB; Limited youth focus; Context is smart homes, not specifically SVAs |
| Adult [15] | Legal Analysis | ATT | Compliance-centric; Lacks youth perspective; No actionable PSE or PPB design levers |
| Adult [16] | Conceptual Analysis | ATT | Lacks youth perspective; No empirical focus on PPB or PSE; Focus on system architecture rather than user behavior or perceptions |

| Population | Method | Construct Coverage | Reported Gaps |
|---|---|---|---|
| Mixed [17] | Theoretical Analysis | ATT | Focuses on system-level transparency and trust, not on measuring users' PPB and PPR; Does not address PSE |
| Youth [18] | Literature Review | PSE, ATT | Lacks empirical focus on PPB; Focus is on design process, not on measuring user outcomes like PSE or ATT |
| Youth [19] | Focus Group, Ethnographic | PSE, ATT, PPBf | Lacks empirical focus on PPB; Limited connection to formal privacy constructs (PCM/PbD/PIPEDA); Focus on critical awareness over PPB |
| Children [20] | Surveys, Interviews, | ATT, PSE | Limited focus on PPB; Small sample size; No actionable insights |
| Mixed [21], [22] | Policy Analysis, Guideline Development | N/A | Lacks empirical focus on user constructs (PSE, PPB, ATT, PPBf); Focus is on organizational compliance rather than measuring user outcomes; Lacks youth focus |

## III. METHODOLOGY

### A. *Research Goals and Questions*

We adopted an interpretivist paradigm to elicit situated accounts of how youth perceive and manage privacy in relation to SVAs. The research was guided by a framework of five key constructs: PPR, PPBf, ATT, PSE, and PPB. To conclude this investigation, we pursued the following Research Questions:

- RQ1 (PPR/PPBf): How do youth articulate everyday trade-offs between perceived privacy risks and perceived benefits when using SVAs?
- RQ2 (ATT): Which transparency touchpoints (e.g., history visibility, mic indicators, permission prompts) most influence reported trust and continued use?
- RQ3 (PSE): In what situations do youth feel able or unable to manage SVA privacy, and what supports raise their self-efficacy?
- RQ4 (PPB): What privacy-protective behaviors (or non-adoptions) do youth report, and what triggers or barriers shape those behaviors?
- RQ5 (ATT to PSE to PPB): Where do device/policy control align or misalign with youths' experiences, and how do these alignments affect self-efficacy and protective behaviors?

### B. *Research Design*

This study received ethics approval from the Vancouver Island University Research Ethics Board (VIU-REB). The approval reference number #103597 was given for behavioral/amendment forms, consent forms, focus group scripts, and questionnaires. Participants were recruited through an initial survey. The survey was distributed through multiple channels, including flyers, emails, personal networks, LinkedIn and through collaboration with several Vancouver Island school districts and Canadian universities to reach our targeted demographic of youth ages 16-24. Participation was entirely voluntary. An incentive was offered to the first 500 survey respondents, with district-specific exceptions where required. The participants had to read and accept a consent form before starting the questionnaire. By submitting the consent form, participants were indicating they understood the conditions of participation in the study as outlined in the consent form. We conducted online surveys through Microsoft Forms. Upon completing the questionnaire, participants were directed to a separate form to claim the incentive by providing their email address. This form also invited participants to indicate if they were interested in being contacted for a focus group.

After contacting those who consented, we conducted 6 focus groups. We use the following naming convention for the qualitative responses: we refer to focus group participants as [FG#-Q#], where FG# denotes the focus group number (1-6), and Q# refers to a unique, anonymized quote identifier from that session (e.g., [FG4-Q2]). Due to a technical difficulty, the third focus group session (FG3) could not be transcribed, resulting in 5 focus groups with 26 participants being available for analysis (no participant data from the lost session is included). Before the focus group, all participants were provided with a consent form to review and accept. Focus groups were conducted and transcribed using Microsoft Teams, with participants instructed to keep their videos off to ensure anonymity. A monetary incentive was provided to all participants who participated in a focus group.

Each of the five constructs (PPR, PPBf, ATT, PSE, and PPB) was operationalized through three open-ended questions, which were used to facilitate discussion within the focus groups. Participants were assigned to a focus group based on the availability they provided during the sign-up process, with efforts made to ensure approximately equal group sizes.

### C. *Data Handling and Analysis*

The data for this study consists of five transcribed focus group sessions (FG1, FG2, FG4, FG5, FG6) with 26 participants. Audio recordings were transcribed verbatim using Microsoft Teams and then checked against recordings for accuracy. A de-identification process replaced all personal identifiers for participants with an anonymous speaker tag, FG#-P# (e.g., FG1-P2), and FG#-MOD for the moderator. The composition of the analyzed focus groups is summarized in Table II. We employed a hybrid, iterative qualitative approach in our analysis, a method established in qualitative HCI research [12]. This approach begins with a deductive scaffold based on existing theories while allowing for inductive refinement from the data itself, ensuring alignment with known constructs while remaining open to new insights. Deductive code families were anchored to the five constructs outlined in Section A of the methodology: PPR, PPBf, ATT, PSE, and PPB. Within each construct, we applied inductive open coding to develop subcodes directly from the data, capturing context-specific meanings and practices.

TABLE II. FOCUS GROUP COMPOSITION

| Item | Value |
|---|---|
| Focus groups | 5 (FG1, FG2, FG4, FG5, FG6) |
| Total participants | N=26 (ages 16-24) |
| Participants per group | Mean 5.5 (range 2-8) |
| Session duration | Total ≈138 min (each ~30-40 min) |
| Mode | Secure videoconference |
| Moderator | Present in all sessions |
| SVA experience | Mixed (regular, occasional, non-users) |

Subcodes and emergent themes were continuously compared within and across focus groups FG1, FG2, FG4, FG5 and FG6 to refine conceptual boundaries, collapse overlapping categories, and identify divergences and negative cases. In each iteration, we ensured that all subcases were grounded in the data. Analytic memos were maintained contemporaneously to document coding decisions, theme development, and reflexive observations regarding assumptions and unexpected findings.

Periodic code-merge meetings were held to resolve disagreements and stabilize the evolving codebook, with particular emphasis placed on negotiated agreement and conceptual clarity rather than reliability coefficients. To strengthen the rigor of our study, we implemented triangulation [13] by integrating youth narrative with a prior audit trail evidence, peer debriefing, negative case analysis, and a saturation/information power rationale. The audit trail comprised versioned codebooks, decision logs, document analysis records, and memo repositories documenting the progression from raw transcripts to overarching themes. Peer debriefing involved regular, structured sessions among researchers to critically examine interpretations and explore alternative explanations for observed patterns in the data. We conducted a systematic search for disconfirming evidence within each construct to refine claims and boundary conditions, thereby mitigating confirmation bias. The entire study flow process, from participant recruitment through to final analysis and reporting, is summarized in Fig. 1.

## IV. RESULTS

Our analysis identified key themes within the five core constructs (PPR, PPBf, ATT, PSE, PPB), directly addressing our research questions. To present the building blocks of our thematic analysis, Table III provides a snapshot of the codebook, detailing the primary code families and their defining subcodes. The findings are structured around these constructs, with each subsection presenting the themes that emerged within them.

### A. Perceived Privacy Risk (PPR)

*1) Theme A1: Ambient Listening And Uncertain Retention Raise Baseline Risk.* Participants consistently expressed anxiety that SVAs are constantly listening to detect wake-words, leading to a perception that conversations could be recorded even during routine use. This concern extended to uncertainty about where data is stored and whether collection persists when devices appear to be “off”. As one participant noted, “I think of the fact that they’re always listening in order to be voice activated” [FG2-Q1], reflecting ambient listening anxiety and how wake-word design is interpreted as continuous capture. A different participant added that “Siri might still [be] listening when… you don’t click the mic icon” [FG6-Q1], reinforcing concerns that devices monitor even outside explicit activation. Another highlights retention fears, stating, “I worry about what happens to the recording of your voice and whether that’s saved somewhere” [FG2-Q2], which inflates perceived harm. Another participant’s suspicion, “How do you know that if you don’t enable them by hand that they’re not still recording you?” [FG2-Q3], signals low trust in software controls. These findings imply a need for clear hardware recording/idle indicators, plain-language retention windows, visible deletion confirmations, and initial setup explanations of how voice processing works.

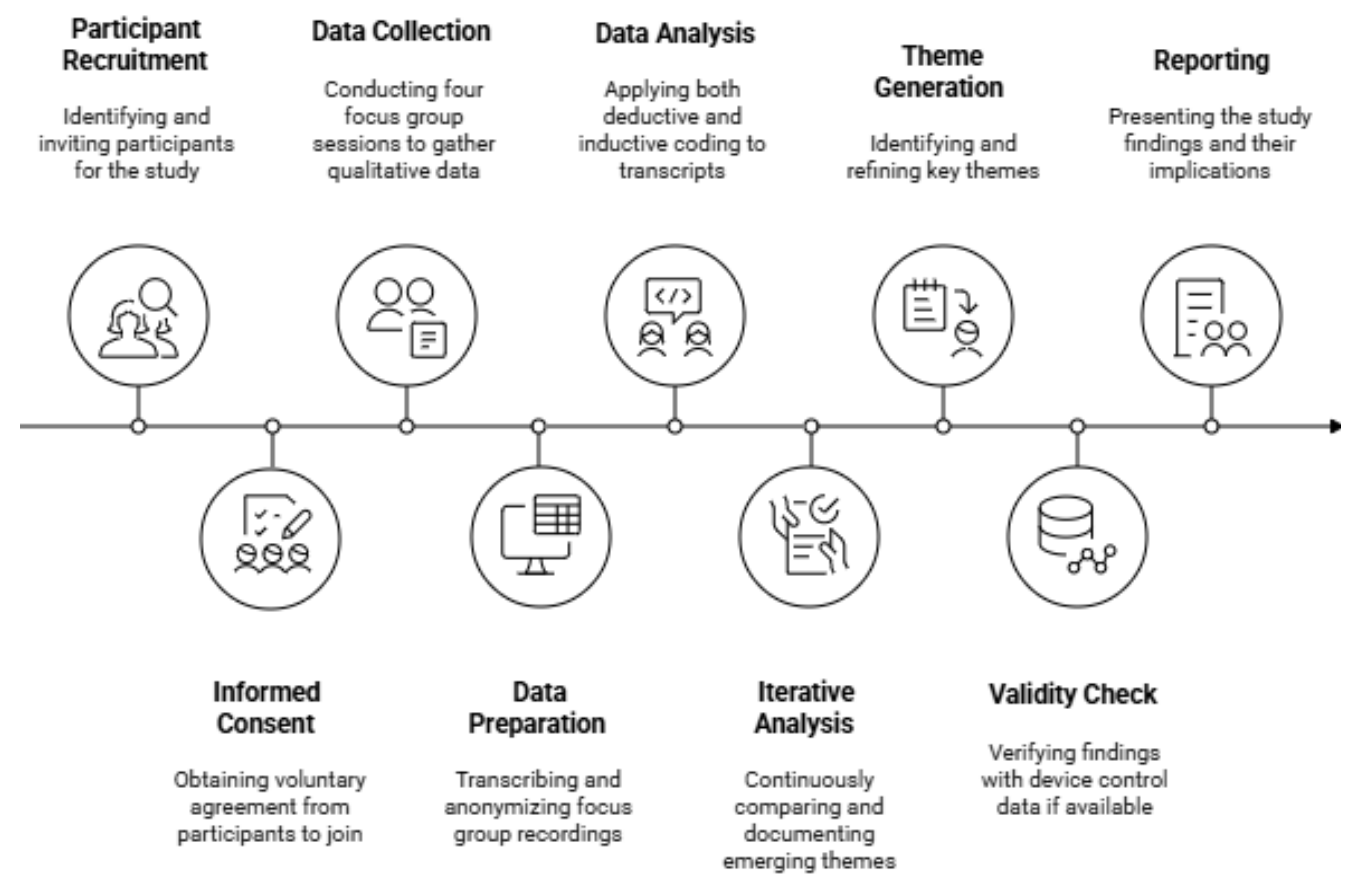

Fig. 1. Study Flow

*2) Theme A2: Suspected Cross-App Inferences Amplify Surveillance Concerns.* Young people also described seeing related content or advertisements after voice interactions, which they interpreted as evidence that their conversations were shared across platforms. One participant remarked, “And after like an hour, I opened Facebook market*place* and suddenly all the rackets showed up for some reason.” [FG*4*-Q1]. This perceived voice-to-ads linkage strengthens the sense of being monitored. A similar sentiment appeared when another participant explained, “Whenever I talk about anything medical related with my parents, I don't like to do it around my phone.” [FG6-Q2], indicating heightened sensitivity around health-related topics. However, not all participants viewed these as serious, as one stated, “I generally don’t really care how they’re using my data and I don’t use any measures” [FG2-Q5] (negative case), illustrating a conscious trade-off of privacy for convenience. This suggests a need for data provenance dashboards that explain recommendations and default settings that separate voice data from advertising ecosystems, allowing personalization only through explicit, reversible opt-in mechanisms to maintain user agency even among those less concerned about privacy.

TABLE III. CODEBOOK SNAPSHOT

| Family | Subcode | Definition | Include | Exclude |
|---|---|---|---|---|
| **PPR** | Ambient “always-listening” | Anxiety that SVAs passively capture to await wake words | Mentions of mics “always on” | Accuracy complaints without a privacy angle |
| **PPR** | Retention unknowns | Uncertainty about storage/length/secondary use | Deletion timelines, account deletion doubts | Non-data bugs |
| **PPBf** | Micro-task convenience | Fast reminders/timers/simple queries | Hands-busy routines | Complex tasks |
| **PPBf** | Entertainment | Music/playback convenience | Always-on entertainment use | Non-voice entertainment |
| **ATT** | Policy overload | Long, unreadable notices | Calls for simpler text | Legal debate without usability |
| **ATT** | Hidden controls | Hard-to-find privacy settings | Discoverability/jargon issues | Pure tap-count complaints |
| **PSE** | Low navigation efficacy | Inability to find/use controls | “Don’t know where to push” | One-off ignorance |
| **PSE** | Device-conditional efficacy | Confidence varies by device | “Phone yes; speaker no” | Global statements |
| **PPB** | Permission refusal | Denying mic/location; uninstall | Keeping prompts on “No” | Complaints without action |
| **PPB** | Physical mitigations | Hardware mute/unplug/separate | Sensor off-switches | Software-only changes |

### B. *Perceived Privacy Benefits (PPBf)*

*1) Theme B1: Micro-Task Convenience Sustains Everyday Use.* Participants primarily value SVAs for simple, time-saving routines. One participant shared, “I like to use it to set reminders because it’s quick and easy.” [FG2-Q6], emphasizing efficiency for daily tasks. Others reported limited use, such as only for timers, indicating a reliance on assistants for narrow purposes. This suggests the importance of keeping micro-tasks accessible with minimal permission and providing just-in-time prompts explaining why specific data is needed. A counterpoint (negative case) emerges from a participant who stated, “I’d rather type the information than be recorded all the time.” [FG4-Q2], revealing that for some, perceived risk still outweighs convenience. Additional evidence comes from a participant who noted, “When I’m studying, I like having Alexa set a timer… If I don’t have my phone, I can just do a timer.” [FG6-Q3], underscoring the contextual value of hands-busy convenience.

*2) Theme B2: Situational Utility in Hands-Busy Contexts And Entertainment Sustain Use.* Voice assistants were highly valued in situations where manual input was impractical. As one explained, “It’s really helpful while you make a call by just speaking to it while your hands might be busy doing some other work...” [FG5-Q1], framing voice use as a tool for safety and accessibility. Entertainment was another key driver, with one participant noting, “I also use it to play music all the time with Alexa.” [FG2-Q8]. These benefits suggest developers should maintain context-aware prompts that support use without requesting broader permissions and offer ‘lite’ entertainment modes with strict privacy defaults to preserve utility.

### C. *Algorithmic Transparency and Trust (ATT)*

*1) Theme C1: Policy Overload And Hidden Controls Undermine Transparency.* Participants found privacy policies difficult to interpret and settings difficult to navigate. One states, “I feel like it’s in the terms and conditions, but you’re not going to read that huge list.” [FG2-Q9], capturing widespread policy fatigue that renders consent meaningless. Others highlighted usability barriers: “I struggle with the settings app.” [FG1-Q1] and “It was too in depth, like under many settings usually need to go or search it online and see how you can find the options.” [FG4-Q3]. This fragmentation drives users off the platform for help. While a minority expressed persistent trust despite opacity (negative case), stating, “I don’t necessarily not trust them to use my information correctly when selling my information.” [FG2-Q10]. Another focus group reinforced this opacity concern, with one participant stating, “They have privacy policies, but nobody reads them… they want to keep that stuff hidden.” [FG6-Q4]. These findings highlight the need for plain-language “data nutrition labels” and a unified privacy hub with searchable, task-based shortcuts.

*2) Theme C2: Retention/Deletion Opacity Depresses Trust.* Unclear deletion lifecycle management, particularly around deletion, created persistent uncertainty. One participant states, “I worry about what happens to the recording of your voice and whether that’s saved somewhere” [FG2-Q2], demonstrating eroding confidence in data governance. To address this, interfaces should implement time-boxed retention defaults (e.g., auto delete after 30-90 days), post-deletion receipts, and a simple activity or audit trail to allow users to verify that their privacy actions are effective.

### D. *Privacy Self-Efficacy (PSE)*

*1) Theme D1: Low Navigation Efficacy Blocks Protective Action.* Many youth participants reported little to no confidence in operating privacy controls. One admitted, “0% - because I don’t know how to.” [FG1-Q2], while another added “ I haven’t done it because I don’t know where to push or what to do in settings” [FG1-Q3]. These statements reveal that usability failures, rather than apathy, are the primary barrier to protective behavior. This suggests interfaces should embed inline micro-tutorials and interactive walkthroughs directly within privacy settings.

*2) Theme D2: Efficacy Is Device-Conditional; Youth Ask For Brief Scaffolds.* Confidence in managing privacy varied sharply by device and platform. Some participants expressed competence only in familiar ecosystems, stating, "For my own device, I feel pretty confident." [FG2-Q13], whereas others confessed total uncertainty: "I don't even know how to use Alexa on my phone" [FG2-Q12]; and "I don't have any confidence in my phone at all." [FG2-Q14]. A sense of hopelessness was also evident, with one participant stating, in regards to devices always "listening", "I feel like they're going to do it, like, no matter what" [FG2-Q15] (boundary case), indicating some users feel any protective action is futile. Another participant added uncertainty about hardware controls: "For my Alexa there's a mute button, but I'm not really certain if that works… I'm not really sure if it's still listening to me." [FG6-Q5]. To strengthen users' sense of control, developers should provide device-specific, 30-second tutorials and clearly expose hardware mute options, explaining their effect on data capture.

### E. Privacy-Protective Behavior (PPB)

*1) Theme E1: Permission And Scope Management Are Primary Mitigations.* Participants reported limiting data collection by refusing permission or uninstalling apps. One stated, "I close microphone permission prompts; I keep it on 'No'." [FG2-Q16], demonstrating a default denial stance. Another reported, "it kept asking me for like permissions, permissions, permissions. So I got a little concerned about that. I just deleted the whole app after that." [FG4-Q4], illustrating active rejection of perceived overreach. Pre-emptive avoidance was also common, such as enabling assistants altogether, as reported by one participant, "Like since I got my new phone, I haven't even enabled like Alexa on it." [FG2-Q12]. However, effort fatigue can constrain sustained action, as captured by "I don't have the energy to track everything or take every safety precaution." [FG2-Q17] (boundary case). An explicit counterpoint arises where participants feel compelled to accept permissions: "For important apps, especially school ones, I need… I just have to accept [permissions]." [FG6-Q7]. These findings highlight the value of granular, revocable permissions, clear request rationales, and pre-setup privacy checklists with conservative defaults.

*2) Theme E2: Physical and Situational Strategies Supplement Software Controls.* Beyond digital settings, youth employed tangible measures for peace of mind. Common actions included disabling sensors: "I just turn off my location and my microphone." [FG5-Q2]. More extreme measures were also reported, such as "I have my phone mic disconnected internally, then if I need to make a call or something, I just use Bluetooth." [FG2-Q18], representing a physical modification driven by profound mistrusts and a willingness to sacrifice functionality for control. Another participant echoed a hardware-first strategy: "I just simply unplug my Alexa or Google Assistant pod." [FG6-Q6]. These findings suggest that devices should include prominent hardware mic-mute switches with clear status indicators. This search for reliable, verifiable interventions implies that devices should include prominent hardware mic-mute switches with clear status indicators and context-aware reminders to review settings after updates.

## V. Discussion

### A. Cross-Construct Mechanism

Our data suggest a qualitative pathway in which friction in ATT, notably policy overload, fragmented settings, and unclear data lifecycles, reduces PSE, which in turn weakens PPB (Theme C1 to D1 to E1). Fig. 2 illustrates this pathway. When youth encounter unreadable policies, multi-hop settings, ambiguous hardware controls, and uncertainty about ambient listening, they report difficulty acting even when motivated, producing "efficacy bottlenecks" at the point of wayfinding. Conversely, low-friction transparency features, plain-language summaries, a unified privacy hub, and visible deletion receipts are associated with higher PSE and more consistent protective actions, without undermining legitimate PPBf such as micro-task convenience and hands-busy use (Themes B1/B2). Situational benefits sometimes override abstract policy concerns: in hands-busy, studying or entertainment contexts, youth prioritize immediate utility, especially when controls are hard to find or interpret (Themes B1/B2 with D1). Boundary cases demonstrate heterogeneity: some participants maintain baseline trust despite opacity (Theme C1 boundary), others experience "mitigation fatigue," and some feel compelled to accept permissions for essential apps, which dampens sustained PPB (Theme E1 boundary). Overall, ATT friction lowers PSE and weakens PPB, while salient PPBf can locally outweigh PPR; targeted transparency that reduces wayfinding friction can raise PSE and strengthen PPB without eroding the value proposition of SVAs.

### B. Relation to prior work

Existing research in HCI describes privacy calculus as the balancing of perceived risks and benefits, and the privacy paradox, where users' stated concerns fail to translate into actions [6], [24], [25]. Our findings expand upon these theories by identifying a more actionable point of failure: a wayfinding (Theme C1) gap created by a lack of transparency and navigational clarity that precedes behavioral breakdown [26], [27]. Participants in this study frequently understood the importance of privacy but lacked the confidence to locate or apply settings effectively (Theme D1). Where prior work on meaningful consent focused on improving readability or timing, our findings emphasize the need for task-bound micro-explanations that occur exactly where control decisions are made [27], [28]. Research on youth digital literacy and PSE links skill and confidence to safer behavior [24], [29], [30]. Our data (Theme D2) refines this by showing that self-efficacy is device-conditional and can be repaired through ultra-brief, embedded scaffolds such as 30-second micro-tutorials. Furthermore, while skepticism about data retention is known [27], our participants explicitly request verifiable solutions like time-boxed default deletion receipts and transparent audit trails [25]. Ultimately, while convenience overrides risk in certain contexts and some participants display baseline trust despite opacity, these patterns reinforce rather than contradict prior accounts of the privacy paradox. Our primary contribution is the empirically grounded ATT to PSE to PPB pathway, which specifies where and how youth-oriented design can intervene to close the gap between privacy awareness and action.

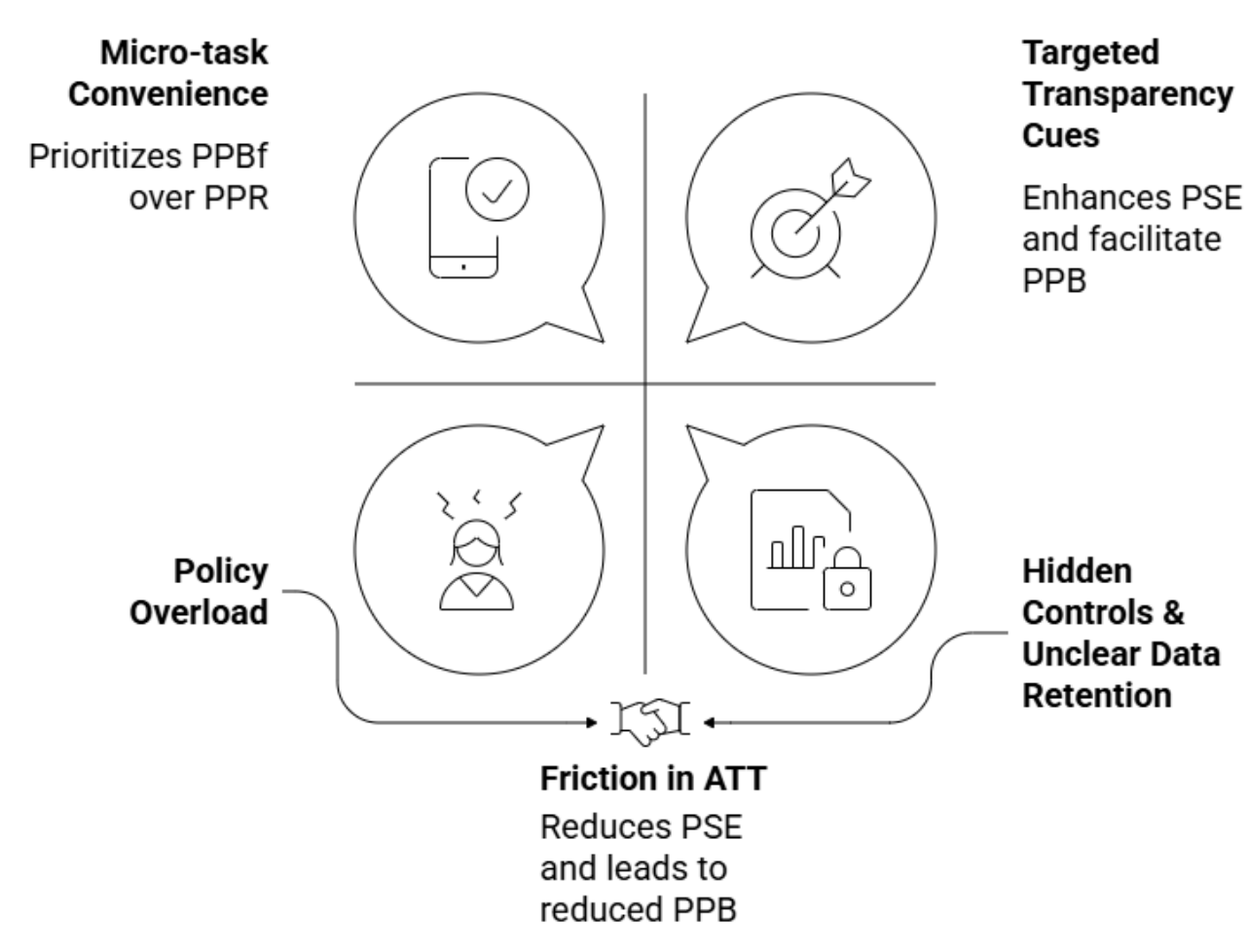

Fig. 2. Construct–Theme Pathway

### C. Design, Policy, and Education Guidance

*1) Design levers (Vendors/Interface Developers)*: Designers should implement a single, unified privacy hub that consolidates all key user actions (review, delete, export data) and provides search with task shortcuts. Controls should feature plain-language "data nutrition labels" explaining what data is collected, why, and for how long, along with just-in-time permission rationales at the moment of request. Persistent microphone-state indicators (both hardware and UI-based) coupled with a one-tap global mute are essential for immediate user control. To build trust, systems should implement time-boxed retention defaults with post-deletion receipts. Embedded, device-conditional micro-tutorials can enhance confidence by guiding users in real-time; a pre-setup privacy checklist with conservative defaults reduces effort during onboarding. Context-aware prompts should support safer use in hands-busy scenarios without escalating permissions, and "lite" entertainment modes should offer strict privacy defaults. Post-update reminders to review permissions help maintain settings over time. These interventions translate transparency into self-efficacy while preserving the convenience that sustains legitimate voice assistant use.

*2) Policy levers (Platforms/Regulators)*: Regulatory frameworks should advance meaningful consent by requiring task-specific micro-notices at the point of interaction rather than relying solely on lengthy privacy policies, and should mandate youth-appropriate retention defaults. Platforms must provide user-visible, auditable logs of data access and deletion events, alongside data provenance disclosures that explain recommendations. To safeguard young users, decoupling voice data from the advertising ecosystems should be the default. Personalization or data sharing with advertisers should occur only through explicit, reversible opt-in consent. Such policy levers align platform accountability with PbD principles suited to youth populations.

*3) Education levers (Schools/Communities)*: Digital literacy initiatives should adopt bite-sized, action-oriented modules that fit youths' attention patterns and daily routines. Examples include 30-second tutorials demonstrating how to review or delete voice history, adjust permissions, and activate hardware mutes. Education should include device-conditional primers that address the self-efficacy gap across different platforms (e.g., phone vs. smart speakers). Finally, scenario-based exercises that let youth practice evaluating privacy trade-offs in realistic contexts can help translate abstract awareness into confident action.

### D. Limitations and Reflexivity

The study captures the youth's perception and self-reported behaviors derived from a limited number of focus groups; therefore, the findings may not generalize across all regions, cultures, or device ecosystems. Recruitment through schools, universities, and personal networks may have introduced selection effects (e.g., higher privacy awareness or technology familiarity). One focus-group session was lost due to a technical failure; its exclusion may have influenced the balance of themes. The videoconference format may have influenced turn-taking and disclosure; however, it also increased accessibility and comfort to participants who might otherwise have been unable or reluctant to attend in person. Self-report introduces recall and social-desirability biases, and we did not collect behavioral logs or telemetry. Perceptions of cross-app "listening" and targeting were not technically verified in this study; we treat them as consequential beliefs that shape behavior rather than as causal claims. We interpret findings in light of prior audit evidence reported in the literature. To ensure rigor, the research team maintained reflexive memos to record assumptions, analytic choices, and interpretive reasoning, while peer debriefing sessions were used to challenge early interpretations. We preserved an audit trail (versioned codebooks, decision logs, memo repositories) and actively searched for disconfirming evidence. The deliberate inclusion of negative cases (e.g., baseline trust despite opacity; mitigation fatigue) helped prevent overgeneralization and increased sensitivity to heterogeneity within youth experiences. Despite the inherent constraints of self-reported data and a modest, Canada-focused sample, the construct-guided framework, constant comparison across groups, and evidence of information power strengthen confidence in the stability of the identified patterns. These results provide a credible foundation for youth-centered SVA design and governance that translates awareness into sustained practice without sacrificing usability.

## VI. Conclusion and Future Work

This research reveals that youth navigation of SVA privacy is defined not by a simple trade-off between risk and benefit, but by a critical transparency self-efficacy pathway. We have detailed how friction in ATT, through policy overload, hidden controls, and opaque data practices, directly undermines PSE, leading to suppressed PPB. The primary contribution of this work is to shift the focus from the existence of the privacy paradox to its mechanism. We identify the "wayfinding gap" as a central failure point and propose concrete, low-friction design interventions, such as unified privacy hubs and verifiable deletion receipts, that can disrupt this cycle by rebuilding self-efficacy, all while preserving the legitimate utility that makes SVAs valuable. Looking forward, future work will build upon this qualitative foundation by co-designing and evaluating these proposed features in situ. Furthermore, we are currently administering a large-scale survey to quantitatively model our proposed pathway, intending to develop a robust, youth-centered framework for SVA privacy that empowers young digital citizens.

## Acknowledgment

This project has been funded by the Office of the Privacy Commissioner of Canada (OPC); the views expressed herein are those of the authors and do not necessarily reflect those of the OPC.